\documentclass[prl,twocolumn,superscriptaddress,floatfix,
 reprint,
preprintnumbers,
nofootinbib,
 amsmath,amssymb,
 aps,
]{revtex4-2}

\usepackage{amsfonts,amsbsy,dsfont,euscript,mathrsfs,fixmath}
\usepackage{amsmath,amssymb,mathtools}
\usepackage{graphicx}
\usepackage{dcolumn}
\usepackage{bm}
\usepackage[bookmarks=true,hyperfigures=true,hidelinks,bookmarksnumbered,unicode]{hyperref}
\usepackage{adjustbox}
\usepackage{multirow}
\usepackage{color, colortbl, xcolor}
\usepackage{bbold}

\usepackage{cancel}
\usepackage{comment}

\newenvironment{eq}
  {\begin{equation}\begin{aligned}}
  {\end{aligned}\end{equation}\ignorespacesafterend}

\DeclareMathOperator{\STr}{STr}

\DeclareMathOperator{\sech}{sech}

\newcommand{\rrf}{q_{\text{\tiny R}}}
\newcommand{\nsf}{q_{\text{\tiny NS}}}
\newcommand{\Jf}{\mathfrak{J}}
\newcommand{\Jc}{\mathcal{J}}
\newcommand{\Rc}{\mathcal{R}}
\newcommand{\x}{\mathsf{x}}
\newcommand{\beq}{\overset{\tts b}{=}}

\newcommand{\tts}[1]{\text{\tiny #1}}

\definecolor{Gray}{gray}{0.9}

\begin{document}

\preprint{HU-EP-26/02-RTG}

\title{Analytic approach to boundary integrability with application to  mixed-flux $AdS_3 \times S^3$ }%

\author{Julio  Cabello Gil}
\email{cabelloj@physik.hu-berlin.de}
\affiliation{Institut f\"ur Theoretische Physik, ETH Z\"urich, Wolfgang-Pauli-Strasse 27, 8093 Z\"urich, Switzerland}
\affiliation{Humboldt-Universität zu Berlin, Zum Großen Windkanal 2, 12489 Berlin, Germany}

\author{Sibylle Driezen}
\email{sdriezen@phys.ethz.ch}
\affiliation{Institut f\"ur Theoretische Physik, ETH Z\"urich, Wolfgang-Pauli-Strasse 27, 8093 Z\"urich, Switzerland}

\date{\today}

\begin{abstract}
Boundary integrability provides rare  analytic control over  field theories with interfaces, from quantum impurity problems to open string dynamics. 
We propose a new approach for integrable boundaries  
in  two-dimensional sigma-models that uses the divisor structure of the  Lax connection to identify admissible reflection maps, including when standard parity-based constructions fail. 
Applied to open strings on $AdS_3\times S^3$ with mixed NSNS-RR flux, where the NSNS flux breaks worldsheet parity, we find two interesting branches of integrable boundaries: one restricted to pure RR flux, and  another  admitting
D-branes wrapping twisted conjugacy classes  for generic flux. 
At the WZW point, the latter reduces to the known conformal D-branes, opening a path to comparison with   conformal perturbation theory. More broadly, our framework   suggests   generalisations of standard lattice constructions that may enlarge existing classifications of integrable boundaries.
\end{abstract}

\maketitle

\setcounter{secnumdepth}{2}

Field theories with interfaces play an important role in both high-energy and condensed-matter physics. In open string theory, worldsheet boundary conditions describe D-branes, while in   condensed-matter and lattice systems they model   boundary critical phenomena  and impurities such as the Kondo effect. When the bulk theory is integrable, the situation becomes especially interesting, as an infinite set of conserved charges  provides a rare source of analytic control for interacting  dynamics. 
The presence of an interface probes this structure in a  nontrivial way. Geometric symmetries are typically broken, and it is far from obvious that an infinite tower of conserved charges can survive. 
When they do, notions of boundary integrability have e.g.~opened access to exact reflection data   \cite{Ghoshal:1993tm} and methods for computing defect observables in AdS/CFT \cite{deLeeuw:2015hxa}.
Identifying which physical interfaces are compatible with integrable dynamics is therefore a basic structural question.
In this work, we address it 
for two-dimensional  sigma-models using  simple analytical arguments from a novel perspective.

The starting point is an integrable sigma-model on  $\mathbb{R}_\tau \times S^1_\sigma$ with an on-shell flat Lax connection ${\cal L}(\x)$,\footnote{For brevity, and where possible, we will suppress explicit time ($\tau \in \mathbb{R}_\tau := \mathbb{R}$) or space ($\sigma \in S^1_\sigma := S^1$ with radius 1) dependence.} i.e. 
\begin{eq} \label{eq:flat-Lax}
\mathrm{d}{\cal L}(\x) + {\cal L}(\x) \wedge {\cal L}(\x) = 0 , \qquad \forall \x \in \mathbb{CP}^1,
\end{eq}
which represents the sigma-model equations of motion
for every value of a spectral parameter $\x$. In general ${\cal L}(\x)$ is a matrix-valued one-form and meromorphic in $\x\in \mathbb{CP}^1$.  Then,  under periodic boundary conditions  $ {\cal L}( \sigma + 2\pi ; \x) =  {\cal L}(\sigma  ; \x) $, the  monodromy matrix $T_{\cal L} (2\pi, 0 ; \x)$, with
\begin{eq} \label{eq:bulk-mono}
T_{\cal L}( b , a ;\x) := \overset{\leftarrow}{\mathcal{P}} \exp \left( - \int^b_a d\sigma  {\cal L}_\sigma (\x) \right), ~~~~ a,b\in \mathbb{R} ,
\end{eq}
is a generator of conserved charges, as its eigenvalues are independent of the choice of spatial slice for every value of $\x$ (see e.g.~the lecture notes \cite{Driezen:2021cpd}).

On an interval $\sigma \in [0,\pi]$, boundary integrability requires a modification of this generator. 
The prototype is obtained by following Sklyanin's construction  \cite{Sklyanin:1987bi,Sklyanin:1988yz}, where one forms a  \textit{double-row} monodromy matrix $T_{\tts b} (\x)$ by gluing $T_{\cal L}( \pi ,0 ; \x)$ over the  interval  $\sigma \in [0,\pi]$ to a reflected copy. On the lattice, 
Sklyanin’s approach applies to integrable  models with  crossing-symmetric $R$-matrices  of difference form \cite{Sklyanin:1988yz}.
The reflection compatible with integrability then has two components: a spatial flip reversing the order of sites,  
and  a spectral reflection $\x \mapsto \rho(\x) = -\x$  implementing  crossing symmetry. 
Additional boundary degrees of freedom are encoded in reflection matrices $K_{\sigma_\star}(\x)$ at the  endpoints $\sigma_\star=0,\pi$, which must satisfy the Boundary Yang-Baxter Equation (BYBE) to preserve integrability. 
In the sigma-model, Sklyanin's equivalent definition of the full reflected copy  can be formulated as
applying a worldsheet parity map $p:(\tau,\sigma)\mapsto(\tau,2\pi-\sigma)$
combined with a spectral reflection $\rho :\mathbb{CP}^1\rightarrow \mathbb{CP}^1$ defined such that the Lax connection transforms covariantly, i.e.~\footnote{Here $p^\ast$ denotes the standard pullback of the   map $p$, which maps the original strip to the reflected copy. That is,  $p^\ast{\cal L}
=
({\cal L}_\tau\circ p)\,\mathrm{d}\tau
-
({\cal L}_\sigma\circ p)\,\mathrm{d}\sigma$.  Similarly, we can rewrite \eqref{eq:par-Lax-inv} as ${p^\ast {\cal L}} = \rho^\ast {\cal L}$, as we will sometimes do below. The purpose of this gluing is to solve fields  on $p(\Sigma)$  in terms of fields on $\Sigma$ for consistent $\rho(\x)$.}
\begin{eq} \label{eq:par-Lax-inv}
p^\ast {\cal L}(\x) = {\cal L}(\rho(\x)), 
\end{eq}
over the full worldsheet $\Sigma$ domain, e.g.~\cite{Sklyanin:1987bi,Mann:2006rh,Dekel:2011ja,Aniceto:2017jor}. Integrable boundary conditions are then obtained by evaluating \eqref{eq:par-Lax-inv} at $\sigma_\star = 0,\pi$, up to a residual gauge freedom by  boundary reflection matrices $K_{\sigma_\star}(\x):=K_{\sigma_\star}(\tau; \x)$.
The corresponding generator of conserved charges is $T_{\tts b} (\x) = K_0 ( \x) T_{\cal L}(2\pi, \pi ; \x ) K_\pi^{-1} (\x) T_{\cal L}(\pi,0; \x)= K_0 ( \x) T_{\cal L}^{-1}(\pi, 0 ; \rho(\x) ) K_\pi^{-1} (\x) T_{\cal L}(\pi,0; \x)$, using \eqref{eq:par-Lax-inv}.\footnote{The original and reflected rows $T_{\cal L}$ together form the non-contractible spatial circle of the doubled construction $[0,\pi] \cup [\pi,2\pi]\sim S^1$, with $K_0$ and $K_\pi$ acting as transition matrices at the gluing points. }

There are however  limitations to taking this  prescription directly as a definition of boundary integrability for sigma-models. 
In fact, there  need  not exist a $\rho(\x)$ that implements \eqref{eq:par-Lax-inv} and thus makes ${\cal L}(\x)$ covariant after a  parity flip. 
This is typically the case when the action contains a parity-odd term, such as a WZ term or a $B$-field, as in many integrable deformations.
In those cases one may consider the lattice equivalent, where $\rho(\x)$ is dictated by the crossing properties of the bulk  $R$-matrix. However such  input is often unavailable (or   indirect through the sigma-model $S$-matrix),  and the defining integrability object is only the  Lax connection ${\cal L}(\x)$. 
There is then no given  principle to select an appropriate generator of conserved charges. 

In this work, the main idea is to address this problem  by allowing a simple generalisation of the reflected copy defined by \eqref{eq:par-Lax-inv}.\footnote{For other alternative  approaches via parity, yet with additional assumptions, see \cite{Bowcock:1995vp,*Corrigan:1996nt}. The approach we take here  moreover admits a natural origin in 4-dimensional Chern-Simons theory,  as will be reported elsewhere \cite{Driezen:2025up}.}  Not only at the boundary points $\sigma_\star$, but also in the definition of the full reflected region can one allow for different gauge representatives. That is, $\rho(\x)$ can be derived as the spectral reflection that consistently solves
\begin{eq}
\label{eq:par-gauge-Lax-inv}
p^\ast {\cal L}(\x) = {\cal L}^U(\rho(\x)) = U {\cal L}(\rho(\x)) U^{-1} - \mathrm{d}U U^{-1}, 
\end{eq}
for all $\tau \in \mathbb{R}$ and $\sigma\in[0,\pi]$, and for some matrix
 $U := U(\tau, \sigma ; \x)$  that is defined over the full worldsheet. 
While a priori a modest change compared to \eqref{eq:par-Lax-inv},  we will see that this extension can lead to very distinct classes of integrable boundary conditions which go beyond those  obtained from parity-covariance alone. 

In   section \ref{sec:reflections}, we give our approach to identify the new admissible gluing data  $(U, \rho(\x))$ by means of the analytic structure of the bulk Lax connection.
In particular, we will explain  how matching
the zero and pole divisors of the two Lax representatives constrains
these possibilities. For each admissible bulk gluing class $(U,\rho(\x))$, the residual
boundary reflection matrices $K_{\sigma_\star}(\x)$ then select the
compatible integrable interfaces through
\eqref{eq:gen-bc-Lax}.

In section \ref{sec:example}, we will illustrate our approach on the distinguished example of the string sigma-model  on $AdS_3\times S^3 \times M_4$ with mixed NSNS and RR flux; for  reviews see \cite{Demulder:2023bux,*Seibold:2024qkh}. 
Its classical Lax connection has been  known for all flux values since the work of Cagnazzo and Zarembo \cite{Cagnazzo:2012se} and illustrates the above limitation: the NSNS term spoils parity-invariance of the action and consequently the parity-based criterion \eqref{eq:par-Lax-inv} alone does not admit a $\rho(\x)$ at nontrivial NSNS flux. Moreover, while worldsheet  methods have led to  $S$-matrices \cite{Hoare:2013ida,*Lloyd:2014bsa,*OhlssonSax:2023qrk,*Frolov:2025uwz,*Frolov:2025tda}, 
a  dual CFT and  lattice formulation 
remains absent for generic fluxes. 
This makes the mixed-flux model an ideal setting to demonstrate  our  approach to boundary integrability.

Boundary integrability  of the mixed-flux  $AdS_3 \times S^3$ background is also inherently motivated by exact worldsheet  studies. At the pure NSNS point, the bosonic theory is described by an exactly solvable  WZW model on  $SU(1,1)\times SU(2)$ \cite{Witten:1983ar,*Maldacena:2000hw}, in which CFT-preserving D-branes wrap twisted conjugacy classes  that admit a boundary CFT (BCFT) description \cite{Alekseev:1998mc,*Stanciu:1999id,*Figueroa-OFarrill:2000gfl,Bachas:2000fr}.  Turning on RR-flux corresponds to a marginal deformation
and, for pointlike branes,  its effect
on the open-string spectrum has been analysed through  boundary conformal perturbation theory, leading to all-orders resummed results, while in the closed-string sector the RR perturbation is singular \cite{Quella:2007sg}. 
The boundary sector thus provides one of the few  controlled windows on mixed-flux physics. It was in fact suggested already in \cite{Cagnazzo:2012se} that identifying integrable boundary conditions could provide  
further structural information at finite flux  and offer a bridge to the BCFT picture. 
The present work realises this at the classical bosonic level. 
Imposing our analytic criteria on the mixed-flux Lax, we find two branches of integrable boundary conditions: one   compatible with the sigma-model action only in the pure RR case, while the other persists for generic flux and includes all D-branes wrapping twisted conjugacy classes. 
Interestingly,  the former arises by identifying $\rho(\x)$ from the covariance of ${\cal L}(\x)$ under parity alone, whereas the latter requires combining  parity with a gauge transformation of the Lax connection.
It is in the second branch that we  have a rigid class of integrable branes whose   embedding  coincides with the conformal WZW branes.  We show that their  worldvolume two-form can remain flux-independent, with the  mixed-flux contribution encoded entirely in dynamical  reflection matrices $K_{\sigma_\star}(\x)$.\footnote{This is compatible with the $\lambda$-deformed WZW case, where integrable D-branes also remain  twisted conjugacy classes but for which
the gauge-invariant worldvolume two-form 
acquires a non-trivial $\lambda$–dependence  given the $\lambda$–deformed metric
\cite{Driezen:2018glg}.}\\

\textit{Outlook}. For  mixed-flux $AdS_3\times S^3$, our results thus  provide a  natural setup to compare integrability-based  data with  conformal perturbation theory around the  solvable WZW point, where the same  twisted conjugacy classes underlie the BCFT description. Including the fermionic sector and analysing the  open string spectrum would enable comparison with the all-order spectra of \cite{Quella:2007sg}
and open the way to extending the analysis beyond the pointlike brane sector. We leave this for future work.

More broadly, our approach may extend to other sigma-models with parity-odd terms, such as integrable deformations, and also suggests a possible generalisation for lattice models.
Traditionally, the BYBE assumes that both rows in the double-row monodromy share the same $R$-matrix. By allowing 
different  representatives  in each row, mirroring different gauge choices of ${\cal L}(\x)$,
a natural question is whether
classifications can be enlarged  
to include
new integrable interfaces  that yield  distinct states after reflection. Given the scarcity of integrable models with interfaces---for  defect observables in holography,   benchmarking impurity physics, engineering  quantum circuits, \textit{etc.}---this could provide valuable new ground for applications of exact methods.

\section{Analytic criteria} \label{sec:reflections} 
To analyse integrable two-dimensional sigma-models with boundaries, let us first collect some necessary background on the    bulk integrable sigma-model. Crucially, there is a Lax connection ${\cal L}(\x)$, generally an object depending locally on the fields and in an $n\times n$ matrix representation of the loop algebra ${\mathbb{C}(\x)}\otimes \mathfrak{g}$, and with the defining property that it  is flat \eqref{eq:flat-Lax} on-shell for all $\x\in\mathbb{CP}^1$. 
The flatness condition \eqref{eq:flat-Lax} can be contrasted with an ordinary current conservation law, $\mathrm{d}\star J = 0$ for a matrix-valued current $J \in \mathfrak{g}$, whose conserved charge is local and obtained by integrating $\star J$ over a spatial slice. At generic $\x$,  the quadratic  term ${\cal L}(\x) \wedge {\cal L}(\x)$ for nonabelian  ${\cal L}(\x)$ then introduces a nonlocality in the corresponding conserved object.
In particular, under periodic  conditions on $S^1_\sigma$, the path-ordered exponential $T_{{\cal L}}( 2\pi , 0 ;\x)$, defined in \eqref{eq:bulk-mono}, is a generating function of generically nonlocal conserved charges.
However, at special points $\x_0$ in $\mathbb{CP}^1$, namely where ${\cal L}(\x_0)=0$, the flatness condition  at first order around $\x=\x_0$  implies that $\star{\cal L}'(\x_0)$ is an ordinary conserved current. The first nontrivial  term of $T_{{\cal L}}( 2\pi , 0 ;\x)$ around $\x_0$ then indeed is a local conserved charge. Hence zeros of the Lax connection in $\mathbb{CP}^1$ hold a special role for the sigma-model physics.

The above statements refer, however, to a particular Lax representative.
In fact, ${\cal L}(\x)$ is not unique: it is defined up to local flatness-preserving gauge transformations
\begin{eq} \label{eq:GGT-Lax}
{\cal L}^U (\x) := U {\cal L}(\x) U^{-1} - \mathrm{d}U U^{-1} , 
\end{eq}
with $U := U(\tau, \sigma; \x)$  an  invertible  matrix that is such that ${\cal L}^U (\x)$ remains in the relevant algebraic realisation. 
Under such gauge transformations one has
\begin{eq}
T_{{\cal L}^U}( 2\pi , 0;\x) = U(\tau, 2\pi; \x) T_{\cal L}( 2\pi , 0 ;\x)  U(\tau, 0; \x)^{-1}. 
\end{eq} 
Thus, for an admissible $U$  that is periodic along the spatial circle, the two monodromies are spectrally equivalent. Their conserved charges are therefore independent of the chosen representative. 
Nevertheless, the spectral points at which their locality is manifest directly from the Lax connection or the monodromy matrix do depend on that representative. Indeed,
$U$ may vary along the spatial slice, and the  inhomogeneous term $\partial_\sigma U U^{-1}$  can  change or  remove the spectral points at
which  ${\cal L}^U_\sigma(\x)$ vanishes. 
This dependence, which does not alter the conserved spectral data of the periodic model,
becomes relevant when different representatives are assigned to the two
rows of the boundary construction, as we will see.

In what follows we will thus denote the gauge orbit of the Lax connection by $[{\cal L}]$, and associate with it the collection
\begin{eq}
\mathcal{Z}([{\cal L}]) := \{ (\x_0, U) ~|~
{\cal L}^{U}(\x_0)=0\}.
\end{eq}
For a particular representative ${\cal L}^U(\x)$, we can denote its zero divisor by
$\mathrm{D}_{\tts z}[{\cal L}^{U}]:=\{ \x_0\in\mathbb{CP}^1 ~|~ {\cal L}^U(\x_0)=0\}$. Similarly we will write $\mathrm{D}_{\tts p}[{\cal L}^{U}]$ for its pole divisor. Note that the condition
${\cal L}^{U}(\x_0)=0$ is equivalent to
${\cal L}(\x_0)=U^{-1}\mathrm{d}U$, so it requires
${\cal L}(\x_0)$ to be locally pure gauge, or equivalently off-shell flat, which is a strong requirement.

Let us now consider open boundaries. As described in the introduction, we approach this by employing a mirror trick, where one glues the original interval $\sigma\in [0,\pi]$ to a reflected, artificial, region $p(\sigma)\in [\pi, 2\pi]$. We then relate the degrees of freedom in the reflected region to the original  ones via the generalised gluing map \eqref{eq:par-gauge-Lax-inv}. The corresponding double-row monodromy matrix then takes the form
\begin{eq} \label{eq:bound-mon}
T_{\tts b}(\x) := K_0 (\x) T^{-1}_{{\cal L}^U}(\pi , 0 ; \rho(\x) ) K_\pi(\x)^{-1} T_{\cal L}(\pi,0; \x) .
\end{eq}
Evaluated along a spatial slice as in   \eqref{eq:bulk-mono}, $T_{\tts b}(\x)$  then generates an infinite tower of conserved charges  under the following \textit{integrable boundary conditions}
\begin{eq} \label{eq:gen-bc-Lax}
{\cal L}_\tau (\x) \beq   K_{\sigma_\star}(\x) {\cal L}_\tau^U (\rho(\x)) K_{\sigma_\star}^{-1}(\x) - \partial_\tau K_{\sigma_\star}(\x) K_{\sigma_\star}^{-1} (\x) ,
\end{eq}
with  ``$\beq$'' denoting evaluation at the boundary $\sigma_\star=0,\pi$, and $K_{\sigma_\star}:=K(\tau;\x)$  an invertible \textit{reflection matrix} defined along   $\partial\Sigma =(\tau ,\sigma_\star=0,\pi)$ and whose action is such that the right-hand-side remains in the relevant algebraic realisation of the Lax connection. These matrices encode
 additional possible boundary degrees of freedom.\footnote{Note furthermore that \eqref{eq:gen-bc-Lax} corresponds precisely to evaluating the time component of \eqref{eq:par-gauge-Lax-inv} at $\sigma_\star$ up to allowing residual gauge transformations by $K_{\sigma_\star}$.} One may  also explicitly verify that $\partial_\tau \mathrm{Tr}T_{\tts b}(\x)^m = 0$ for any integer $m$, using flatness of the Lax and the boundary conditions  \eqref{eq:gen-bc-Lax}, so that expanding the eigenvalues of $T_{\tts b}(\x)$ in $\x$ indeed produces conserved objects. Throughout, we
use boundary integrability in this Lax sense of a conserved double-row
monodromy and its spectral invariants.

Let us emphasize that, although only the restriction of $U$ to the boundary enters
\eqref{eq:gen-bc-Lax}, its role is fundamentally distinct from that of
$K_{\sigma_\star}$.
The matrix $U$ is defined throughout the bulk and identifies the transformation of the Lax under parity, thereby selecting
the Lax representative used in the reflected region. Its relative action between
the two rows is therefore part of the bulk gluing data. In the next subsection, we will fix  the gluing data $(U, \rho(\x))$ from the bulk analytic structure,
prior to solving for the  boundary data $K_{\sigma_\star}$.
This viewpoint generalises   the prescription of \cite{Sklyanin:1987bi,Mann:2006rh,Dekel:2011ja,Aniceto:2017jor,Gombor:2018ppd,Driezen:2018glg,Driezen:2019ykp} beyond  parity-covariant gluing, as described in the introduction.

\textit{Remark:} 
The gauge transformation matrices do not need to belong to the Lie group $G$ of $\mathfrak{g}$, yet their conjugation should preserve $\mathfrak{g}$. Constant matrices, e.g. $K_{\sigma_\star} = w$, can thus implement (outer)
automorphisms, $W(y)=wyw^{-1}\in\mathfrak{g}$ for $y\in \mathfrak{g}$, as in \cite{Dekel:2011ja}.
In suitable realisations,
such matrices also accommodate 
the  common alternative convention for $T_{\cal L}$ in the reflected region, where the inverse in \eqref{eq:bound-mon} is replaced by transposition.

\paragraph{\textbf{Fixing  $U$ and $\rho(\x)$ from bulk analyticity.}} Our main point is  that $U$ and $\rho(\x)$ can be fixed  from  the analytic structure of ${\cal L}(\x)$, mirroring the lattice logic  where $\rho(\x)$ is derived from bulk properties of the $R$-matrix (rather than boundary $K$-matrices).

A first useful criterion is to choose ``physical''  representatives for the unreflected  and reflected rows, such that $T_{\tts b}(\x)$ can generate a set of local charges. 
Concretely, we thus select two elements
of the gauge orbit that  realise nontrivial elements in ${\cal Z}([{\cal L}])$. 
Without loss of generality, we can assume that the given representative for ${\cal L}$ is already calibrated to do so, and thus we take $(\x_0, \mathbb{1}), (\x_0', U)\in{\cal Z}([{\cal L}])$, i.e.~with
\begin{eq}
\label{eq:lax-res-zero}
{\cal L}(\x_0)=0,
\qquad
{\cal L}^{U}(\x'_0)=0 .
\end{eq}
We then let ${\cal L}$ be  the reference representative in the unreflected row on $\sigma\in [0,\pi]$, and ${\cal L}^U$ the representative for the reflected row.\footnote{Note  that the double-row generator $T_{\tts b}(\x)$ built from representatives ${\cal L}^{U_+}$ and ${\cal L}^{U_-}$ is spectrally equivalent to the one built from ${\cal L}$ and ${\cal L}^{U=U_+^{-1}U_-}$ upon the boundary transformation $K_{\sigma_\star} \mapsto U_+^{-1}(\sigma_\star) K_{\sigma_\star} U_+(\sigma_\star)$. Thus only the relative choice of Lax representative  is
meaningful, as a common gauge transformation leaves the spectrum of $T_{\tts b}(\x)$ 
unchanged. \label{f:spectral-equiv}}

In practice,  the set ${\cal Z}([\cal L])$, and thus all relevant gauges to construct $T_{\tts b}(\x)$ under our requirements,  is  found by scanning for those points $\x\in \mathbb{CP}^1$ where  ${\cal L}(\x)$ becomes locally pure gauge.
Relaxing these criteria for $T_{\tts b}(\x)$ would formally enlarge the possibilities of finding integrable boundary conditions, but could      obscure the existence of residual local symmetries which are usually desired for spectral analyses. 

Given relevant representatives ${\cal L}$ and ${\cal L}^U$, the spectral reflection $\rho(\x)$ is then constrained directly by the bulk gluing map
\eqref{eq:par-gauge-Lax-inv}. Since this map  must return to the original row when applied
twice, we require that $\rho : \mathbb{CP}^1 \rightarrow \mathbb{CP}^1$ is  a nontrivial involutive Möbius transformation
\begin{eq} \label{eq:inv-rho}
\rho(\x) = \frac{a \x +b}{c \x - a} ,  \quad a^2 + bc \neq 0 ,
\end{eq}
with $\rho^2(\x)=\x$ and $a,b,c \in \mathbb{C}$  to be determined by consistency. In fact, 
because \eqref{eq:par-gauge-Lax-inv} is an equality of meromorphic
Lax one-forms,
their complete zero and pole divisors must be mapped into one
another, i.e.
\begin{eq} \label{eq:div-for-rho}
\mathrm{D}_{\tts p}[{\cal L}]=\mathrm{D}_{\tts p}[\rho^\ast{\cal L}^U] , 
\qquad \mathrm{D}_{\tts z}[{\cal L}]=\mathrm{D}_{\tts z}[\rho^\ast{\cal L}^U ]  .
\end{eq} 
The zeros selected above then obey
$\rho(\x_0)=\x_0'$. This allows local charges to be extracted from the leading terms of the expansion of $T_{\tts b} (\x)$  around $\x_0$.\footnote{Note that $K_{\sigma_\star}$ is $\sigma$-independent and thus does not alter locality (assuming $K_{\sigma_\star}$ is regular at $\x_0$ and is itself local). }

Once the relevant Lax representatives have been classified modulo constant
gauge transformations,  an involutive Möbius transformation can only fix marked points or
exchange them in pairs while preserving their analytic type. For a finite set of such representatives, as in the example discussed below, this leaves only finitely many
candidates $(U,\rho)$ modulo constant  gauge transformations. Each choice of $(U,\rho)$, together with all subsequent  $K_{\sigma_\star}$ solutions of \eqref{eq:gen-bc-Lax}, can lead to different integrable boundary conditions and hence preserve different
combinations of the bulk local charges.

Let us end by clarifying the scope of this prescription. Within the class of admissible local representatives for which both rows possess a zero, every solution of the bulk gluing map \eqref{eq:par-gauge-Lax-inv} must satisfy the
analytic matching conditions \eqref{eq:div-for-rho}  above. These conditions are necessary but
not sufficient, so the resulting candidates $(U,\rho)$ must still be checked
against the full gluing map. In fact,  consistency under two applications additionally requires the resulting composite gauge transformation to act trivially on the Lax connection, as it does in the
examples below. In addition, the integrable boundary conditions \eqref{eq:gen-bc-Lax} must be consistent with the sigma-model variational problem (cf.~Appendix \ref{app:sigma-bcs}). Furthermore, the requirement of a zero in the row representatives can be relaxed, which may enlarge the space of solutions. The corresponding double-row monodromies may then
still generate conserved spectral invariants, but  an
expansion producing local conserved charges would no longer apply. 
Our
prescription is thus intended to capture the class relevant for 
local analyses, rather than all possible reflected Lax pairs.

\section{Mixed-flux $AdS_3\times S^3$} \label{sec:example}

We now apply our approach to boundary integrability in the mixed-flux $AdS_3\times S^3$ coset sigma-model \cite{Cagnazzo:2012se}, and analyse the simplest resulting integrable boundary conditions. The relevant background ingredients for this analysis are reviewed in App.~\ref{app:sigma-bcs} and \ref{app:ads3s3}, to which we refer the reader for conventions and notation.

For simplicity, we restrict to the bosonic sector ($g_{\tts F}=\tilde{g}_{\tts F}=\mathbb{1}$) of the mixed-flux $AdS_3\times S^3$ coset sigma-model action \eqref{eq:S-action} and fix   the coset-gauge $\tilde{g}_{\tts B} =\mathbb{1}$, so that  \eqref{eq:bosonic-gauge-simp} applies. 
The Lax connection \eqref{eq:bulk-lax-gen} then simplifies to
\begin{eq}
    {\cal L}(\x) = f_1 (\x)\Jc^{\tts{B}} +f_2(\x) \star \Jc^{\tts{B}} ,
\end{eq}
with
\begin{eq}
f_1(\x) &=\frac{ \rrf   }{\x^2-1}  + \frac{\rrf+1}{2}, ~~   f_2(\x) = \frac{  \x \rrf   }{1-\x^2} +\frac{\nsf}{2} . 
\end{eq}
where  $\rrf, \nsf \in \mathbb{R}$ denote the amount of RR and NSNS flux of the full theory respectively, which are related as $\rrf^2+\nsf^2=1$ \cite{Cagnazzo:2012se}.

\paragraph{\textbf{Marked points, spectral reflections and  boundary conditions.}} 
There are two interesting Lax representatives in the gauge orbit of ${\cal L}(\x)$ that realise nontrivial elements in ${\cal Z}([{\cal L}])$. In fact, for ${\cal L}(\x)$ to be pure gauge, a necessary requirement is $f_2(\x)=0$, which has the following two solutions
\begin{eq}
    \x_0 = \frac{\rrf -1}{\nsf} , \qquad \x_0' = \frac{\rrf +1}{\nsf} .
\end{eq}
Upon $\rrf^2+\nsf^2=1$ one then finds\footnote{When keeping the coset-gauge for $\tilde{g}_{\tts B}$ general,  ${\cal L}(\x)$ would be pure gauge only at $\x=\x_0'$, with $U=g$. In the pure RR case this point reduces to $\x=\infty$. This corresponds to the standard set-up for coset models where one expands ${\cal L}^{g}(\x)$ to obtain local charges.  
The possibility of having an additional pure gauge location  when using $\tilde{g}_{\tts B} =\mathbb{1}$ (i.e.~at $\x=\x_0$ with $U=\mathbb{1}$) is however special to  $AdS_3\times S^3$ (permutation-)cosets.  }
\begin{eq}
    {\cal L}(\x_0) = 0 , \qquad {\cal L}(\x_0') = g_{\tts B}^{-1} \mathrm{d}g_{\tts B} .
\end{eq}
Hence, modulo constant gauge transformations, the two relevant elements of
${\cal Z}([{\cal L}])$ are represented by
\begin{eq}
    (\x_0,\mathbb{1}),
    \qquad
    (\x_0',g_{\tts B}) .
\end{eq}
Both ${\cal L}(\x)$ and ${\cal L}^{g_{\tts B}}(\x)$ are thus admissible representatives for each row in $T_{\tts b}(\x)$.

Both ${\cal L}(\x)$ and ${\cal L}^{g_{\tts B}}(\x)$ have the same pole divisor:  two simple poles at $\x_\pm= \pm 1$. 
Given \eqref{eq:div-for-rho}, there are then four possible reflections $\rho_i(\x)$, depending on whether the poles and zeros are fixed or permuted. 
The behaviour of the zeros  further determines the  necessary gauges for the Lax connections: when $\rho(\x_0)=\x_0$ we must take $U=\mathbb{1}$,  while for $\rho(\x_0)=\x_0'$ we must take $U=g_{\tts B}$.\footnote{Note that  $\rho(\x_0')=\x_0'$ with \textit{both} Lax representatives in the gauge $U=g_{\tts B}$ is equivalent to  $\rho(\x_0)=\x_0$ with $U=\mathbb{1}$ in both rows up to redefining $K_{\sigma_{\star}}$, as explained in footnote \ref{f:spectral-equiv}.}
In  table \ref{tab:involutions}, we summarise the possibilities,  the resulting solutions of \eqref{eq:inv-rho} and the resulting  gluing maps \eqref{eq:par-gauge-Lax-inv}. The latter are equivalent at the boundary to the integrable boundary conditions \eqref{eq:gen-bc-Lax} for trivial $K_{\sigma_{\star}}$ matrices. 
\begin{table}[h]
\centering
\begin{tabular}{|c|c|c|c|c|}
\hline
 $i$ & Zeros  & Poles  &   $\rho_i(\x)$ &  
  gluing maps  \\ 
\hline
1 & $ \rho_1(\x_0)= \x_0$ & $\rho_1(\x_\pm) =  \x_\mp$  & $\tfrac{-\x-\nsf}{\nsf \x+1}, ~ \nsf\neq 1$ & 
$p^\ast{\cal L} = \rho^\ast{\cal L}$ 
\\
2 & $\rho_2(\x_0) = \x_0'$ & $\rho_2(\x_\pm) =  \x_\mp$ &  $-\x^{-1}$ & 
$p^\ast{\cal L} = \rho^\ast{\cal L}^{g_{\tts B}}$ 
\\
3 & $\rho_3(\x_0)= \x_0$ & $\rho_3(\x_\pm )= \x_\pm$   & $\x^{-1}, ~~ \rrf = 0$ &  
$p^\ast{\cal L} = \rho^\ast{\cal L}$  \\
4 & $\rho_4(\x_0) = \x_0'$ & $\rho_4(\x_\pm )= \x_\pm$ &  $\x^{-1}, ~~\nsf=0$ & 
$p^\ast{\cal L} = \rho^\ast{\cal L}^{g_{\tts B}}$  \\
\hline
\end{tabular}
\caption{The inequivalent Möbius involutions $\rho(\x)$. 
}
\label{tab:involutions}
\end{table}

Let us first consider Cases 3 and 4, in which the poles are fixed. Case 3, defined additionally  by $\rho_3(\x_0)=\x_0$, is formally solved by $\rho_3(\x)=\x^{-1}$  only at $\rrf=0$. 
At this point, however, the apparent zero $\x_0=-1$ collides with a pole, while the pole residues  vanish and the bosonic Lax connection becomes independent of $\x$. We therefore discard this degenerate case.\footnote{Moreover, for trivial $K_{\sigma_{\star}}$ one   has $T_{\tts b}(\x)=\mathbb{1}$, while \eqref{eq:gen-bc-Lax} becomes an identity and therefore supplies no boundary conditions to be matched with the
sigma-model variational conditions \eqref{eq:sm-bcs}. 
}
Case 4, defined  by $\rho_4(\x_0)=\x_0'$ and thus with $U=g_{\tts B}$, is  solved by $\rho_4(\x)=\x^{-1}$  only at $\nsf=0$. 
In this pure-RR limit, we have $\x_0=0$ and
$\x_0'=\infty$, and the spectral reflection is in principle admissible.
Suppose then  that $K_{\sigma_\star}(\x)$ is regular and nondegenerate at
$\x_\pm$. Defining $W_\pm(y)
:=
K_{\sigma_\star}(\x_\pm)\,
y\,
K_{\sigma_\star}^{-1}(\x_\pm)$, the residue conditions following from \eqref{eq:gen-bc-Lax} at $\x_\pm$  consequently are $(1+W_\pm\operatorname{Ad}_{g_{\tts B}})
\Jc^{\tts B}_\mp\beq0$. Whenever these two operators are invertible---as is generally the case at generic target-space points---the conditions imply, after
translating to the $AdS_3\times S^3$ fields as in \eqref{eq:cur-to-X} and using coordinates $\sigma^\pm = \tau \pm \sigma$, that
\begin{eq}
\partial_+X^\mu\beq0,
\qquad
\partial_-X^\mu\beq0 , \qquad \forall~\mu =1,\cdots, 6 .
\end{eq}
They  constrain both the temporal and spatial derivatives of every field at the
boundary and are therefore overdetermined and incompatible with the
sigma-model variational conditions \eqref{eq:sm-bcs}.\footnote{This conclusion may be weakened on exceptional loci where either operator
develops a kernel, or for reflection matrices that are degenerate or singular
at $\x_\pm$ and can modify the pole structure. We do not analyse these
possibilities here.} 

More  interesting cases arise when  poles are permuted: 
\textit{Branch 1.} Firstly, we have 
\begin{eq} \label{eq:rho-1}
\rho_1(\x) = -\frac{\x + \nsf}{\nsf \x +1} , \qquad  \nsf \neq 1 ,
\end{eq}
which fixes $\x_0$ and thus takes $U=\mathbb{1}$. For $\nsf=0$, this reduces to the standard reflection $\rho_1(\x)=-\x$ obtained from parity-covariant gluing \eqref{eq:par-Lax-inv} in pure RR cosets \cite{Dekel:2011ja,Linardopoulos:2021rfq,*Linardopoulos:2022wol,*Demjaha:2025axy}. For general $\nsf$,  mapping the residues of the  integrable boundary conditions \eqref{eq:gen-bc-Lax} for trivial $K_{\sigma_\star}$, i.e.~${\cal L}_\tau (\x) \beq {\cal L}_\tau (\rho_1(\x))$, now  gives 
\begin{eq} \label{eq:branch1-rel}
\Jc_+^{\tts B} \beq \frac{1-\nsf}{1+\nsf} \Jc^{\tts B}_- ~~ \Rightarrow ~~ \partial_\sigma X^\mu \beq -\nsf \partial_\tau X^\mu ,
\end{eq}
where we again used coordinates $\sigma^\pm = \tau \pm \sigma$ and translated  to the $AdS_3\times S^3$ fields $X^\mu$ as in \eqref{eq:cur-to-X}. In this diagonal form, the above conditions can only be compatible with the sigma-model boundary conditions \eqref{eq:sm-bcs} at $\nsf=0$, i.e.~\textit{only in the pure RR case}. Indeed,  $\nsf\neq0$ 
 would require
${\cal F}_{\mu\mu}\neq 0$, which is impossible because
${\cal F}_{\mu\nu}$ must be antisymmetric. Consequently, for $\nsf=0$ the relation \eqref{eq:branch1-rel} implies that the integrable D-brane is space-filling   ($p=5$) with  a vanishing worldvolume flux ${\cal F}_{mn}=0$. Allowing more general  matrices  $K_{\sigma_\star}$ can however change  the dimension $p$ and the flux ${\cal F}$. For example, for a constant $K_{\sigma_\star}=w$ implementing an automorphism $W$, eq.~\eqref{eq:gen-bc-Lax} additionally implies $W^2=\mathbb{1}$, such that $W$ must have eigenvalues $\lambda_\pm=\pm 1$ of multiplicity $m_\pm$ respectively, with $m_+ + m_- =\mathrm{dim}(AdS_3\times S^3)=6$. Going to a diagonal frame $X^\mu \rightarrow \tilde{X}^\mu (X)$ and demanding compatibility with  \eqref{eq:sm-bcs} gives again valid   conditions  only at $\nsf=0$,   given by
\begin{eq}
\partial_\sigma \tilde{X}^{\mu_+} \beq 0 , \qquad \partial_\tau \tilde{X}^{\mu_-} \beq 0 ,
\end{eq}
where $\mu_\pm$ has cardinality $m_\pm$. For the   automorphism given in \eqref{eq:W-out}, one finds e.g.~$m_+=4$ and then this corresponds to a D-brane with $p=3$ and ${\cal F}=0$. Nontrivial worldvolume fluxes are typically introduced by dynamical and $\x$-dependent $K_{\sigma_\star}$-matrices, as happens for several interesting D-branes in $AdS_5\times S^5$ {and $AdS_4 \times \mathbb{CP}^3$} \cite{Linardopoulos:2021rfq,*Linardopoulos:2022wol,*Demjaha:2025axy,Linardopoulos:2025ypq}.
In the future,  it would be interesting to compare with the $\x$-dependent quantum $K_{\sigma_\star}$-matrices of \cite{Prinsloo:2015apa,*Bielli:2024xuv,*Bielli:2025abu} known for pure RR flux. For nontrivial mixed-flux, an interesting question is also whether compatibility between \eqref{eq:rho-1} and \eqref{eq:sm-bcs}    requires dynamical  reflection data $K_{\sigma_\star}$.

\textit{Branch 2.} Both poles and zeros are permuted by
\begin{eq}
\rho_2(\x) = -\x^{-1} ,
\end{eq}
with no  restriction on $\nsf,\rrf$. In this branch we have $U=g_{\tts B}$. Including  automorphisms $W$ as above and mapping residues,  the  integrable boundary conditions \eqref{eq:gen-bc-Lax} for trivial $K_{\sigma_\star}$, i.e.~${\cal L}_\tau (\x) \beq W\left[{\cal L}^{g_{\tts B}}_\tau (\rho_2(\x)) \right]$, now simplify  for $\rrf\neq 0$ to
\begin{eq} \label{eq:int-conf-min}
\Jc^{\tts B}_\pm \beq - W\left[\Rc^{\tts B}_\mp \right] = - (W\mathrm{Ad}_{g_{\tts B}}) \left[ \Jc^{\tts B}_\mp \right] ,
\end{eq}
implying the constraint $(W\mathrm{Ad}_{g_{\tts B}})^2 \beq \mathbb{1}$. Interestingly, for $\rrf=0$, where ${\cal L}(\x)$ and ${\cal L}^{g_{\tts B}}(\x)$ become $\x$-independent, the boundary conditions enhance to $\Jc^{\tts B}_+ \beq - W\left[ \Rc_-^{\tts B} \right]$ without further constraints. These are precisely the conformal boundary conditions of WZW models  which correspond to D-branes embedded as twisted conjugacy classes in $SU(1,1) \times SU(2)$.\footnote{In the WZW limit, integrability is instead encoded by the Kac-Moody algebra.} For $W=\mathbb{1}$  this yields \textit{unstable} $dS_2 \times S^2$ branes \cite{Bachas:2000fr}, which we will exclude in the following, 
while including the nontrivial outer automorphism \eqref{eq:w-out} with $K_{\sigma_\star}=w_{\tts o}$ yields \textit{stable} $AdS_2\times S^2$ branes \cite{Bachas:2000fr}. In terms of the fields $X^\mu = \{\psi, \omega , t ; \alpha, \beta, \gamma\}$ given in \eqref{eq:gB-params}, the latter boundary conditions read
\begin{equation} \label{eq:wzw-bcs}
\begin{alignedat}{3}
\partial_\tau \psi &\beq  0 , \qquad &&\partial_\sigma \omega  &&\beq \tanh\psi\cosh\omega \partial_\tau t , \\ &  &&\partial_\sigma t  &&\beq  \tanh\psi\sech \omega \partial_\tau \omega , \\
\partial_\tau \alpha &\beq  0 , \qquad &&\partial_\sigma \beta  &&\beq - \cot\alpha \sin\beta \partial_\tau \gamma , \\ & &&\partial_\sigma \gamma  &&\beq  \cot\alpha \csc\beta \partial_\tau \beta .
\end{alignedat}
\end{equation}
The conjugacy classes are thus fixed at $\{\psi ; \alpha\}$ and extend along $X^m=\{t, \omega ; \beta , \gamma\}$. With \eqref{eq:sm-bcs} and \eqref{eq:met-nsns} the worldvolume flux can then be extracted from \eqref{eq:wzw-bcs} as
\begin{eq} \label{eq:wzw-flux}
{\cal F} = \tfrac{\sinh2\psi \cosh\omega}{2} \mathrm{d}\omega \wedge \mathrm{d}t + \tfrac{\sin2\alpha\sin\beta}{2} \mathrm{d}\gamma \wedge \mathrm{d}\beta .
\end{eq}
For mixed-flux, however, the constraint 
$(W\mathrm{Ad}_{g_{\tts B}})^2 \beq \mathbb{1}$  fixes the D-brane conjugacy-classes on specific loci, i.e.~at
\begin{eq} \label{eq:loci}
\psi = 0 , \qquad \alpha = \{ 0 , \tfrac{\pi}{2}, \pi \} ,
\end{eq}
on which the worldvolume flux becomes trivial, ${\cal F}=0$. 
We next show that this restriction can be lifted by allowing nontrivial reflection data (i.e.~beyond $K_{\sigma_\star}=w_{\tts o}$).

Before extending Branch 2, 
let us emphasise the  distinction between
Branch 1 and 2. While Branch 2 is admissible at generic mixed flux, including pure RR flux,
Branch 1 gives rise to additional integrable boundary conditions only in the
pure-RR limit.  This has a simple origin in the $\mathbb{Z}_2$ symmetries of the bulk sigma-model action \eqref{eq:S-action} already visible in the bosonic sector. At pure RR flux, $\nsf=0$, the bosonic action is invariant under worldsheet parity, $\sigma\rightarrow2\pi-\sigma$. Adding the WZ term however  breaks this symmetry. Yet for generic mixed flux  the action is invariant under  a combination of worldsheet parity and group element inversion:
\begin{eq}
    \sigma\rightarrow2\pi-\sigma,
    \qquad
    g_{\tts B}\rightarrow g_{\tts B}^{-1}.
\end{eq}
These two symmetries underlie the spectral reflections \eqref{eq:par-gauge-Lax-inv} associated with Branches 1 and 2, respectively, and explain why the pure-RR case enjoys more types of    integrable boundary conditions.
It would be interesting to relate this to  crossing properties of \cite{Hoare:2013ida,*Lloyd:2014bsa,*OhlssonSax:2023qrk,*Frolov:2025uwz,*Frolov:2025tda}.

\paragraph{\textbf{Extension to all twisted conjugacy classes.}}
To restore the full family of stable $AdS_2\times S^2$ twisted conjugacy classes with two-form \eqref{eq:wzw-flux} at generic $(\psi,\alpha)$, we allow for appropriate  matrices $K_{\sigma_\star}(\x)$ while keeping the same  reflection $\rho_2(\x)=-\x^{-1}$.
For this purpose we reverse the logic and use the WZW boundary conditions \eqref{eq:wzw-bcs}  as input to \eqref{eq:gen-bc-Lax} to determine the required $K_{\sigma_\star}(\x)$.

We find that it is sufficient to consider the ansatz
\begin{eq}\label{eq:dynU}
K_{\sigma_{\star}} (\tau ; \x) = \left( \mathbb{1} +  \sum_{k=1}^6 u_k (X^\mu  ; \x)    T_k 
\right) w_{\tts o}   \ ,
\end{eq}
with $T_k$ the $\mathfrak{su}(1,1)\oplus\mathfrak{su}(2)$ generators given in \eqref{eq:generators} and $u_k$ depending on $\tau$ only through the target-space fields $X^\mu  =\{\psi,\omega,t;\alpha,\beta,\gamma\}$.
Projecting the resulting system \eqref{eq:gen-bc-Lax} onto the  generators yields  six coupled first-order PDEs  in the worldsheet variables $(\tau , \sigma)$. 
Using \eqref{eq:wzw-bcs}, we rewrite them in terms of $\partial_\sigma\{\psi,\alpha\}$ and $\partial_\tau\{t,\omega,\beta,\gamma\}$ only.
Requiring that  $u_k$ does not depend on $\partial_\sigma \{\psi, \alpha\}$ (i.e.~$u_k$ depends on the fields but not   their derivatives), gives a set of algebraic conditions solved by
\begin{equation}
\begin{alignedat}{3}
&\tfrac{u_2 (\tau ; \x)}{u_1 (\tau ; \x)} =  \cot t  , \quad &&\tfrac{u_3 (\tau ; \x)}{u_1 (\tau ; \x)} =  \tanh\omega \csc t  , \\
&\tfrac{u_5 (\tau ; \x)}{u_4 (\tau ; \x)} = -\cot \gamma  , \qquad &&\tfrac{u_6 (\tau ; \x)}{u_4 (\tau ; \x)}= - \cot \beta \csc \gamma   .
\end{alignedat}
\end{equation}
While this reduces the unfixed functions to $u_1$ and $u_4$ only, we remarkably find that the resulting six PDEs can still  be solved consistently. We simply obtain
\begin{eq}
u_1 = -\sqrt{2} \frac{\sinh 2\psi \cosh\omega \sin t}{z( \x) + \cosh2\psi} , ~~ u_4 = \frac{\sqrt{2}}{i} \frac{\sin 2\alpha \sin\beta \sin\gamma}{z(\x) - \cos 2\alpha} ,
\end{eq}
with the rational  function $z(\x)$ defined as
\begin{eq}
z(\x) = \frac{(\x-1)(\nsf (\x+1) +\rrf (\x-1) ) }{(\x+1)((1-\nsf) (\x-1) + \rrf (\x+1))}   .
\end{eq}
The resulting solution $K_{\sigma_\star}$ is regular and nondegenerate at $\x_\pm$.
On the loci \eqref{eq:loci} we obtain $u_k=0$ and hence $K_{\sigma_\star}=w_{\tts o}$, as well as 
whenever the denominator of $z(\x)$ vanishes, 
which occurs at $\x=\x_0$ and at $\rrf=0$ (the WZW point). This is therefore fully consistent with the previous subsection.  Away from these special points,
we  find that branes wrapping twisted conjugacy classes, as in boundary WZW models, remain compatible with integrability, 
with the mixed-flux deformation carried entirely by the nontrivial reflection data $K_{\sigma_\star}(\x)$ rather than the brane geometry.

\vspace{2pt}

We will report on the extension to the fermionic sector, including the analysis of residual symmetries from $T_{\tts b}(\x)$ and the  open-string spectrum,  elsewhere.

\vspace{10pt}

\paragraph{\bf Acknowledgements.}
We thank  D.~Bielli, C.~Kristjansen, S.~Lacroix, G.~Linardopoulos, B.~Nairz,  D.~Polvara, A.~Retore, and K.~Zarembo for useful discussions, and  A.~Retore, K.~Zarembo and especially S.~Lacroix for   comments on the draft. 
This work is partly based
on the Master thesis of JCG  prepared at ETH Z\"urich and we thank N.~Beisert for acting as co-advisor. JCG's research is funded by the Deutsche Forschungsgemeinschaft (DFG, German Research Foundation) - Projektnummer 417533893/GRK2575 ``Rethinking Quantum Field Theory''.  SD is supported by the Swiss National Science Foundation through the SPF fellowship
TMPFP2$\_224600$ and the NCCR SwissMAP. SD also thanks the organisers and participants of the workshop \textit{Higher-$d$ integrability} (Favignana, 2025) where part of this work was presented. 

\setcounter{secnumdepth}{2}
\appendix

\section{Sigma-model boundary conditions}\label{app:sigma-bcs}

We formulate  integrable boundary conditions through a relation on the Lax connection~\eqref{eq:gen-bc-Lax}, which guarantees conservation of charges generated by $T_{\tts b}(\x)$. At the level of the sigma-model,  this must translate into  boundary conditions on the fields $X^\mu$, $\mu=0,\ldots, D-1$, that are compatible with the variational principle. 

The bosonic action of a standard two-derivative  sigma-model on $\Sigma=\mathbb{R}_\tau \times [0,\pi]_\sigma$ can be written as \footnote{Here we use the flat sigma-model metric $\eta_{\alpha\beta}=\mathrm{diag}(1,-1)$, with $\sigma^\alpha = (\tau, \sigma)$,  and $\epsilon_{\tau\sigma}=1$. In general the target space fields (metric and B-field) are curved, 
$G_{\mu\nu}:= G_{\mu\nu}(X)$ and $B_{\mu\nu}:= B_{\mu\nu}(X)$
and $T$ is the sigma-model coupling (i.e.~string tension). As an action of open strings \eqref{eq:sm-action} is written in conformal gauge.  }
\begin{eq}\label{eq:sm-action}
S
={}& \frac{T}{2}\int_\Sigma \mathrm{d}^2\sigma\;
\partial_\alpha X^\mu
\big(\eta^{\alpha\beta}G_{\mu\nu}+\epsilon^{\alpha\beta}B_{\mu\nu}\big)
\partial_\beta X^\nu \\
&+\int_{\partial\Sigma}\mathrm{d}\tau\; A_m\,\partial_\tau X^m  ,
\end{eq}
where $m, n=0, \cdots, p \leq D-1$ are the directions along which $X^\mu\vert_{\partial \Sigma}$ is allowed to fluctuate, i.e.~(for open strings) those along a D$p$-brane with a $U(1)$ gauge field $A_m := A_m(X)$  and field strength $F_{mn} = \partial_m A_n - \partial_n A_m$.
The boundary equations of motion are then solved by
\begin{equation}
\begin{alignedat}{2} \label{eq:sm-bcs}
&\text{Dirichlet:} \quad &&\partial_\tau X^{\hat{m}} \beq 0 , \\
&\text{gen.~Neumann:} \quad && G_{mn} \partial_\sigma X^n \beq {\cal F}_{mn}  \partial_\tau X^n ,  
\end{alignedat}
\end{equation}
with $\hat{m} = p+1 , \cdots , D-1$ and ${\cal F}_{mn} (X) = B_{mn}(X) + T^{-1} F_{mn}(X)$ the gauge-invariant worldvolume flux. 
Here we assumed adapted coordinates in which the target-space fields split orthogonally in the  Neumann  ($m$) and Dirichlet directions ($\hat{m}$).
Note that the 
conditions \eqref{eq:sm-bcs} are rather restrictive,  given the symmetry properties of each term in the generalised Neumann conditions. 

The problem addressed in this work can  be restated as determining those Dirichlet and Neumann conditions  which are guaranteed to preserve classical integrability. 
For the  sigma-models considered here,
the Lax connection   ${\cal L}(\x) \in \mathbb{C}(\x) \otimes \mathfrak{g}$ is a linear combination of  (projections of) $\mathfrak{g}$-valued one-form currents ${\cal J}$ and $\star {\cal J}$
with $\x$-dependent coefficients. 
The currents themselves are linear in $\mathrm{d} X^\mu$ so that one can  introduce a frame field $E$ as
\begin{eq} \label{eq:cur-to-X}
\Jc_\alpha = E_\mu{}^A (X) T_A \partial_\alpha X^\mu , \quad  (\star \Jc)_\alpha =   \Jc_\beta \epsilon^\beta{}_\alpha  ,
\end{eq}
with $T_A$ the generators of $\mathfrak{g}$.
The conditions  \eqref{eq:gen-bc-Lax} for  chosen $(\rho,U,K_{\sigma_\star})$ 
therefore yield linear gluing relations between
$\partial_\sigma X^\mu$ and $\partial_\tau X^\mu$ at $\partial\Sigma$
which must be matched to the expressions \eqref{eq:sm-bcs} for consistency with the variational principle, and hence  determine 
the dimension $p$ and the flux $\mathcal{F}_{mn}$ (equivalently $F_{mn}$ for given $B_{mn}$).  
Likewise, one may start from local D-brane data $(p,{\cal F})$ which determine the relations \eqref{eq:sm-bcs}, and solve eqs.~\eqref{eq:par-gauge-Lax-inv} and \eqref{eq:gen-bc-Lax} for $(\rho,U,K_{\sigma_\star})$. 

\section{Mixed-flux $AdS_3\times S^3$ bulk sigma-model} \label{app:ads3s3}
In this appendix, we collect the background ingredients for the boundary integrability analysis of the mixed NSNS/RR flux string on $AdS_3\times S^3 \times M_4$, with $M_4$  a four-dimensional compact manifold required to complete the model to a type IIB superstring (see \cite{Demulder:2023bux,*Seibold:2024qkh} for reviews). Since the integrable structure relevant for our purposes is encoded in the $AdS_3\times S^3$ sector, we will restrict attention to its Green-Schwarz formulation as a $\mathbb{Z}_4$-graded supercoset sigma-model \cite{Cagnazzo:2012se} (see also \cite{Babichenko:2014yaa,*Hoare:2014oua}), while keeping $M_4$ implicit throughout.
\paragraph{\textbf{Supercoset.}}
Concretely, the worldsheet degrees of freedom in this sector are described by  $\widehat{F} / F^{(0)}$ with
\begin{eq}
\widehat{F} =G \times G  , \qquad G=PSU(1,1|2) ,
\end{eq}
and $F^{(0)}$ the bosonic diagonal subgroup of $\widehat{F}$, i.e.~$(h,h)\in F^{(0)}$ for $h\in SU(1,1)\times SU(2)$.
The Lie superalgebra $\hat{\mathfrak{f}} = \mathfrak{g} \oplus \mathfrak{g}$, with $\mathfrak{g}=\mathfrak{psu}(1,1|2)$,  of $\widehat{F}$ admits a $\mathbb{Z}_4$ grading
\begin{eq}
\hat{\mathfrak{f}} = \mathfrak{f}^{(0)} \oplus \mathfrak{f}^{(1)} \oplus \mathfrak{f}^{(2)} \oplus \mathfrak{f}^{(3)} , 
\end{eq}
with $\mathfrak{f}^{(0)}$ the Lie algebra corresponding to $F^{(0)}$. The bosonic subalgebra $\mathfrak{su}(1,1)\oplus \mathfrak{su}(2)$ is generated by
\begin{equation} \label{eq:generators}
    \begin{gathered}
T_{1,2} = \tfrac{1}{\sqrt{2}} \left( 
\sigma_{1,3} \oplus 0
\right) , \qquad T_{3} = \tfrac{i}{\sqrt{2}} \left( 
\sigma_{2} \oplus 0
\right) , \\
T_{4,5,6} = \tfrac{1}{\sqrt{2}} \left(0 \oplus \sigma_{1,2,3} \right) ,
    \end{gathered}
\end{equation}
with $\sigma_{1,2,3}$ the  Pauli matrices. 
For a complete superalgebra realisation of ${\mathfrak{g}}$ and $\hat{\mathfrak{f}}$, see e.g.~App.~C of \cite{Seibold:2020ouf}.

Let $f(\tau, \sigma) \in \widehat{F}=G\times G$ be a supercoset representative with $\Jf=f^{-1}\mathrm{d}f \in \hat{\mathfrak{f}}$ the left-invariant current. We will  work in a block-diagonal matrix representation, where $f=\mathrm{diag}(g, \tilde{g})$ and $\Jf = \mathrm{diag}(\Jc , \widetilde{\Jc} ) $  with   $g,\tilde{g}\in G$,
and   implement the $\mathbb{Z}_4$-grading  as
\begin{equation}
    \begin{alignedat}{2}
\Jf^{(0)} &= \begin{pmatrix} \Jc^{(0)} & \\ & \Jc^{(0)}  \end{pmatrix}  , \quad 
&&\Jf^{(1)} = \begin{pmatrix} \Jc^{(1)} & \\ & -i\Jc^{(1)}  \end{pmatrix} , \\
\Jf^{(2)} &= \begin{pmatrix} \Jc^{(2)} & \\ & -\Jc^{(2)}  \end{pmatrix}  , \quad 
&&\Jf^{(3)} = \begin{pmatrix} \Jc^{(3)} & \\ & i\Jc^{(3)}  \end{pmatrix}  , 
\end{alignedat}
\end{equation}
where $\Jf^{(i)}\in \mathfrak{f}^{(i)}$ and $\Jc^{(0)} =\tfrac{1}{2} P_{\tts{e}} (\Jc + \widetilde{\Jc})$, $\Jc^{(2)} =\tfrac{1}{2} P_{\tts{e}} (\Jc - \widetilde{\Jc})$, $\Jc^{(1)} = \tfrac{1}{2}P_{\tts{o}} (\Jc + i\widetilde{\Jc})$,  $\Jc^{(3)} =\tfrac{1}{2} P_{\tts{o}} (\Jc -i \widetilde{\Jc})$, with $P_{\tts{e}}$  ($P_{\tts{o}}$) projectors on the Grassmann-even (odd) parts of $\mathfrak{g}$.   We further define two inequivalent  bilinear forms on $\hat{\mathfrak{f}}$ as 
\begin{eq}
\STr (\Jf )= \mathrm{Str} (\Jc) + \mathrm{Str}( \widetilde{\Jc}) , \quad \widetilde{\STr}(\Jf) = \STr (\mathfrak{W} \Jf) ,
\end{eq}
with $\mathrm{Str}$  the  supertrace of $\mathfrak{g}$ and $\mathfrak{W} = \mathrm{diag}(\mathbb{1}, - \mathbb{1} )$. 

\paragraph{\textbf{Action and integrability.}}
The bulk sigma-model action with mixed-flux contributions is \cite{Cagnazzo:2012se,Babichenko:2014yaa,*Hoare:2014oua}
\begin{eq} \label{eq:S-action}
S_{\tts{cl.}} ={}& \frac{1}{2} \int_\Sigma \STr \left( \mathfrak{J}^{(2)} \wedge \star \mathfrak{J}^{(2)} + \rrf \Jf^{(1)} \wedge \Jf^{(3)} \right)  \\
&+   \nsf \int_{\cal B} \widetilde{\STr} \left( \frac{2}{3} \Jf^{(2)} \wedge \Jf^{(2)}\wedge \Jf^{(2)} \right. \\ & + \left. \Jf^{(1)} \wedge \Jf^{(3)}\wedge \Jf^{(2)} + \Jf^{(3)} \wedge \Jf^{(1)}\wedge \Jf^{(2)} \right) , 
\end{eq}
where $\rrf$ and $\nsf$ respectively quantify the RR and NSNS flux couplings and ${\cal B}$ is a three-dimensional manifold such that $\Sigma = \partial {\cal B}$.  Throughout we   will consider the branch $0\leq \nsf \leq 1$. 
The full bosonic $AdS_3 \times S^3 \times T^4$ worldsheet theory is  obtained by reinstating the string tension $T = (2\pi\alpha ')^{-1}$ and  supplementing the above action with four flat bosonic fields, which we keep implicit. 
The pure NSNS background $\rrf=0$, $\nsf =1$ corresponds to the WZW model, i.e.~in the bosonic subsector the $SL(2, R)\times SU(2)$ WZW description of
strings on $AdS_3\times S^3$ \cite{Witten:1983ar,*Maldacena:2000hw}. The usual WZ level $k$ is  related to the above parameters by $2\pi T \nsf = k$
with $k\in \mathbb{Z}$
due to the compact $SU(2)$ isometry subgroup. 

Classical integrability, $\kappa$-symmetry  and   conformal invariance to one-loop in  $\alpha'$ require the following relation between the flux parameters \cite{Cagnazzo:2012se}
\begin{eq} \label{eq:cons-pars}
\rrf^2 + \nsf^2=1 .
\end{eq}
The flat Lax connection then takes the form
\begin{align} 
{\cal L}(\x) ={}& {\cal J}^{(0)}+ \rrf \frac{\mathsf{x}^2+1}{\mathsf{x}^2-1} \Jc^{(2)} + \left( \nsf - \frac{2\rrf \x}{\x^2-1} \right) \star \Jc^{(2)} \nonumber \\
&+ \left( \mathsf{x} + \frac{\rrf}{1- \nsf} \right) \sqrt{\frac{\rrf (1-\nsf)}{\x^2-1}}\Jc^{(1)} \label{eq:bulk-lax-gen} \\
&+ \left( \mathsf{x} - \frac{\rrf}{1+ \nsf} \right) \sqrt{\frac{\rrf (1+\nsf)}{\x^2-1}}\Jc^{(3)} . \nonumber
\end{align}
At the pure RR point $\rrf=1$, $\nsf =0$ this Lax corresponds to \cite{Bena:2003wd}, while at the pure NSNS/WZW point it degenerates and  does not evidently capture the full dynamics. In that case one can instead  work with holomorphic current algebras rather than  Lax connections. 

\paragraph{\textbf{Parametrisation and background.}}
The sigma-model \eqref{eq:sm-action} is invariant under local, right-acting, $F^{(0)}$ gauge transformations. We can therefore take 
\begin{eq} 
g= g_{\tts F} g_{\tts B} , \qquad \tilde{g} = \tilde{g}_{\tts F} \tilde{g}_{\tts B}=\tilde{g}_{\tts F},
\end{eq}
in $f=\mathrm{diag}(g, \tilde{g})$, where $g_{\tts B} \in SU(1,1)\times SU(2)$, $\tilde{g}_{\tts B}=\mathbb{1}$  and $g_{\tts F}, \tilde{g}_{\tts F} \in \exp (P_{\tts o} \mathfrak{g})$. In this case, we have
\begin{eq}
\Jc = \Jc^{\tts{B}}  + g_{\tts B}^{-1} \Jc^{\tts{F}} g_{\tts B} , \qquad \widetilde{\Jc} = 
 \widetilde{\Jc}^{\tts F} ,
\end{eq}
with $\Jc^{\tts B}= g_{\tts B}^{-1} \mathrm{d} g_{\tts B} \in \mathfrak{su}(1,1)\oplus \mathfrak{su}(2)$,   $\Jc^{\tts F}= g_{\tts F}^{-1} \mathrm{d} g_{\tts F} \in \mathfrak{g}$ and $\widetilde{\Jc}^{\tts F}= \tilde{g}_{\tts F}^{-1} \mathrm{d} \tilde{g}_{\tts F} \in \mathfrak{g}$.
In the main text, we furthermore focus on the bosonic sector of 
$\mathfrak{psu}(1,1|2)$ and thus take $g_{\tts F}, \tilde{g}_{\tts F} = \mathbb{1}$. Altogether, this implies the simplifications
\begin{eq} \label{eq:bosonic-gauge-simp}
\left.\Jc^{(0)} \right\vert_{g_{\tts B}} = \left. \Jc^{(2)}\right\vert_{g_{\tts B}} = \frac{1}{2} \Jc^{\tts{B}} , \qquad \left.\Jc^{(1,3)}\right\vert_{g_{\tts B}}=  0 .
\end{eq}
We will then parametrise the bosonic element $g_B =\mathrm{diag}(g_{SU(1,1 ) }~|~g_{SU(2)} ) $ as
\begin{eq} \label{eq:gB-params}
g_{SU(1,1)} &= \sinh\psi~ \sigma_3 + \cosh\psi~ e^{\tfrac{it}{2} \sigma_2} e^{\omega\sigma_1}  e^{\tfrac{it}{2} \sigma_2}  , \\
 g_{SU(2)} &=e^{-\tfrac{i\gamma}{2} \sigma_3}e^{\tfrac{i\beta}{2} \sigma_1} e^{i \alpha \sigma_3}  e^{-\tfrac{i\beta}{2} \sigma_1} e^{\tfrac{i\gamma}{2} \sigma_3} ,
\end{eq}
where  ${\psi, \omega, t} \in \mathbb{R}^3$ are global coordinates on $AdS_3$, with $\{\omega , t\}$ parametrising $AdS_2$ fibred over $\psi$, while the coordinates $\beta \in [0,\pi)$ and $\gamma \in [0,2\pi)$ parametrise an $S^2$ fibered over the interval $\alpha \in [0,\pi)$, forming the three-sphere $S^3$. The metric and NSNS three-form $H=\mathrm{d}B$ can then be expressed  as
\begin{eq} \label{eq:met-nsns}
\mathrm{d}s^2={}& \mathrm{d}\psi^2 + \cosh^2\psi \left(- \cosh^2 \omega \mathrm{d}t^2 + \mathrm{d}\omega^2 \right) \\  &+ \mathrm{d} \alpha^2 + \sin^2\alpha \left(\mathrm{d}\beta^2 + \sin^2\beta\mathrm{d}\gamma^2 \right), \\
H ={}&  2 \nsf \cosh^2 \psi \cosh \omega  \mathrm{d}\psi \wedge \mathrm{d}t \wedge \mathrm{d}\omega \\ &+2 \nsf \sin^2 \alpha \sin \beta \mathrm{d}\alpha \wedge \mathrm{d}\beta \wedge \mathrm{d}\gamma . 
\end{eq}
When including the fermionic sector, the RR 3-form flux  is $F_3 = e^{-\Phi_0} \rrf (H/\nsf)$ for constant dilaton $\Phi(X) = \Phi_0$. 

\paragraph{\textbf{Outer automorphisms of $\boldsymbol{SU(1,1)\times SU(2)}$.}} Automorphisms $W$ of $\mathfrak{su}(1,1) \oplus \mathfrak{su}(2)$ which preserve the ad-invariant metric form the group $SO(2,1) \times SO(3)$. Explicit matrix realisations  can be defined as $W(y) = w y w^{-1} = y^A W_A{}^B T_B$ for $y=y^AT_A\in\mathfrak{g}$. The elements relevant to the boundary integrability discussion of the main text are those $W$ which are involutive, $W^2=\mathbb{1}$.   Up to conjugation (and  combinations), the ones that do not mix $\mathfrak{su}(1,1)$ and $\mathfrak{su}(2)$ and that \textit{(i)} are inner are
\begin{eq}
W_{\tts{inn}} = 
\left\{\begin{array}{ll}
\mathbb{1}_3 \oplus \mathbb{1}_3 \\
\mathrm{diag}(-1,-1,1) \oplus \mathbb{1}_3  \\
 \mathbb{1}_3 \oplus \mathrm{diag}(-1,-1,1) 
\end{array}\right.  ~\in SO^+(2,1)\times SO(3) ,
\end{eq}
 or \textit{(ii)} are outer are
\begin{eq} \label{eq:W-out}
W_{\tts{out.}}  = \mathrm{diag}(-1,1,-1) \oplus \mathbb{1}_3 \in SO^-(2,1)\times SO(3) . 
\end{eq}
The latter is realised as $W_{\tts{out.}}(X) =w_{\tts o}X w_{\tts o}^{-1}$ with
\begin{eq} \label{eq:w-out}
w_{\tts o} = \mathrm{diag}(1,-1) \oplus \mathbb{1}_2  \not\in SU(1,1)\times SU(2).
\end{eq}


\bibliography{apssamp}

\end{document}